# Skin Lesion Analyser: An Efficient Seven-Way Multi-Class Skin Cancer Classification Using MobileNet

Saket S. Chaturvedi [1] [0000-0003-0700-404X] *, Kajol Gupta[1], and Prakash S. Prasad[1]

[1] Department of Computer Science & Engineering, Priyadarshini Institute of Engineering & Technology, Nagpur-440019, India.
Email: saketschaturvedi@gmail.com

"This is a pre-copyedited version of a contribution published in Advances in Intelligent Systems and Computing, Hassanien A., Bhatnagar R., Darwish A. (eds) published by Chaturvedi S.S., Gupta K., Prasad P.S. The definitive authentication version is available online via https://doi.org/10.1007/978-981-15-3383-9_15 ".

**Abstract.** Skin cancer is an emerging global health problem with 123,000 melanoma and 30,00,000 non-melanoma cases worldwide each year. The recent studies have reported excessive exposure to ultraviolet rays as a major factor in developing skin cancer. The most effective solution to control the death rate for skin cancer is a timely diagnosis of skin lesions as the five-year survival rate for melanoma patients is 99 percent when diagnosed and screened at the early stage. Considering an inability of dermatologists for accurate diagnosis of skin cancer, there is a need to develop an automated efficient system for the diagnosis of skin cancer. This study explores an efficient automated method for skin cancer classification with better evaluation metrics as compared to previous studies or expert dermatologists. We utilized a MobileNet model pretrained on approximately 12,80,000 images from 2014 ImageNet Challenge and finetuned on 10015 dermoscopy images of HAM10000 dataset employing transfer learning. The model used in this study achieved an overall accuracy of 83.1% for seven classes in the dataset, whereas top2 and top3 accuracy of 91.36% and 95.34%, respectively. Also, the weighted average of precision, weighted average of recall, and weighted average of f1-score were found to be 89%, 83%, and 83%, respectively. This method has the potential to assist dermatology specialists in decision making at critical stages. We have deployed our deep learning model at https://saketchaturvedi.github.io as web application.

**Keywords:** Skin Cancer, Dermoscopy, Classification, Convolutional Neural Network.

## 1 Introduction

Skin Cancer is an emerging global health problem considering the increasing prevalence of harmful ultraviolet rays in the earth's environment. The researchers had discovered a further 10 percent depletion of the ozone layer will intensify the problem of skin cancer with an additional 300,000 non-melanoma and 4,500 melanoma cases each year [1]. Currently, every year 123,000 melanomas and 30,00,000 non-melanoma cases are recorded worldwide [1]- [5]. The recent study on the prevention of skin cancer reports 90 percent of non-melanoma and 86 percent of melanoma cases induced by excessive exposure of ultraviolet rays [6], [7]. The UV radiation detriments the DNA present at the inner layers of skin, triggering the uncontrolled growth of skin cells, which may even emerge as a skin cancer [8].

The most straightforward and effective solution to control the mortality rate for skin cancer is the timely diagnosis of skin cancer as the survival rate for melanoma patients in a five-year timespan is 99 percent when diagnosed and screened at the early stage [9], [10]. Moreover, the most mundane skin cancer types BCC and SCC are highly treatable when early diagnosed and treated adequately [9], [11]. Dermatologist primarily utilizes visual inspection to diagnose skin cancer, which is a challenging task considering the visual similarity among skin cancers. However, dermoscopy has been popular for the diagnosis of skin cancer recently considering the ability of dermoscopy to accurately visualize the skin lesions not discernible with the naked eye. Reports on the diagnostic accuracy of clinical dermatologists have claimed 80 percent diagnostic accuracy for a dermatologist with experience greater than ten years, whereas the dermatologists with experience of 3-5 years were able to achieve diagnostic accuracy of only 62 percent, the accuracy further dropped for less experienced dermatologists [12]. The studies on Dermoscopy imply a need to develop an automated efficient, and robust system for the diagnosis of skin cancer since the fledgling dermatologists may deteriorate the diagnostic accuracy of skin lesions [13]–[16].

Although the method is complicated, deep learning algorithms have shown exceptional performance in visual tasks and even outperformed humans in gaming, e.g., Go [17], Atari [18] and object recognition [19], which has lead to conduct the research on automated screening of skin cancers [9]. Several studies have been done to compare the dermatologist-level, and Deep Learning based automated classification of skin cancer [20], [21]. Esteva et al. reported a benchmark study comparing the performance of dermatologists and a CNN model over 129,450 clinical images, showing the CNN model performs at par or better than dermatologists [21]. In recent years, the trend has shifted to Deep Neural Networks (DNNs) [22], which were proposed to overcome the drawbacks of previous models [9], [23]–[30]. Although DNNs require huge data for the training, they have an appealing impact on medical image classification [30]. The current literature mostly employs transfer learning to solve large dataset problem. Transfer Learning is a method where a model trained over another similar task is finetuned for the given task. Mostly, the melanoma screening works employing DNNs have trained a network from scratch [26], [31], or employs transfer knowledge [24], [25], [27], [28] from ImageNet. The main difference between them the DNN architecture and implementation framework — Caffe [24], [27] is the most common framework, and ResNet [28], AlexNet [25], VGG-16 [29] are most common architectures.

Previous work in dermoscopic automated skin cancer classification has lacked generality capability [30], [32], [33], and have not achieved pleasing results for multi-class skin cancer classification [21], [34], [36]. This study explores an efficient automated method for the classification of dermoscopy skin cancer images. We utilized a MobileNet convolutional neural network [37] pretrained on approximately 12,80,000 images from 2014 ImageNet Challenge [31] and finetuned on HAM10000 dataset [38] which contain 10015 dermoscopy images employing transfer learning [39]. The MobileNet model classified skin lesion image with performance better or comparable to expert dermatologists for seven classes. We also conducted data-analysis on the dermoscopy images of skin cancer from HAM10000 dataset to uncover the relation of skin cancer with several parameters to strengthen the understanding of skin cancer.

## 2 Method

### 2.1 Dataset

We have utilized HAM10000 Dataset [38] for the training and validation in this study. HAM10000 dataset is a benchmark dataset with over 50% of lesions confirmed by pathology. The dataset consists of a total of 10015 dermoscopy images, which includes 6705 Melanocytic nevi images, 1113 Melanoma images, 1099 Benign keratosis images, 514 Basal cell carcinoma images, 327 Actinic keratosis images, 142 Vascular images and 115 Dermatofibroma images with 600X450 pixels resolution. Sample images of skin cancer types from HAM10000 are represented in Figure 1.

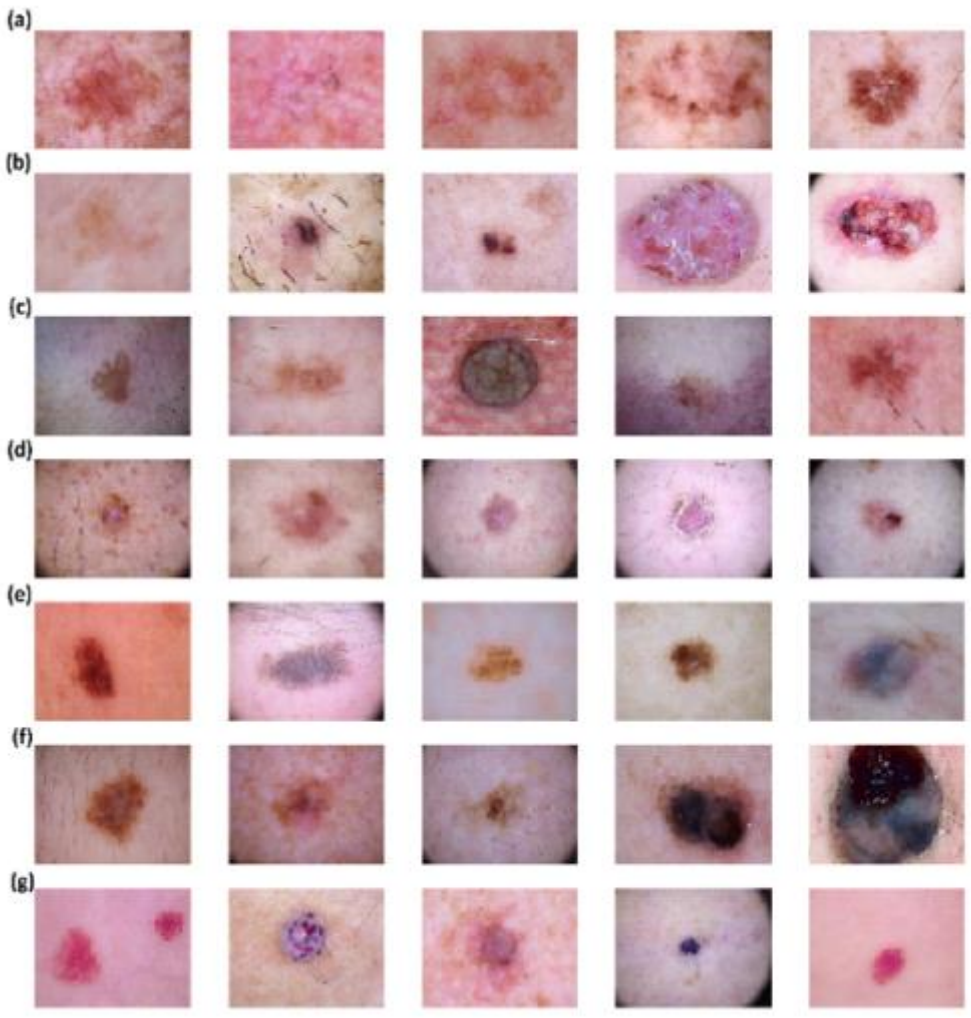

**Fig. 1.** Sample images from HAM10000 dataset for cancer types (a) Actinic Keratosis (b) Basal Cell Carcinoma (c) Benign Keratosis (d) Dermatofibroma (e) Melanocytic nevi (f) Melanoma (g) Vascular Lesions

### 2.2 Data pre-processing

The pre-processing of skin lesion images was done by using Keras ImageDataGenerator [40]. The 57 null Age entries in the dataset were filled using the mean filling method [41]. The Dermoscopy images in the dataset were downscaled to 224X224 pixel resolution from 600X450 pixel resolution to make images compatible with the MobileNet model [37]. The 10015 images in the dataset were split into the training set (9077 images) and validation set (938 images). The dataset images with no duplication in training data were selected for the validation set so that the authenticity in the validation process can be maintained.

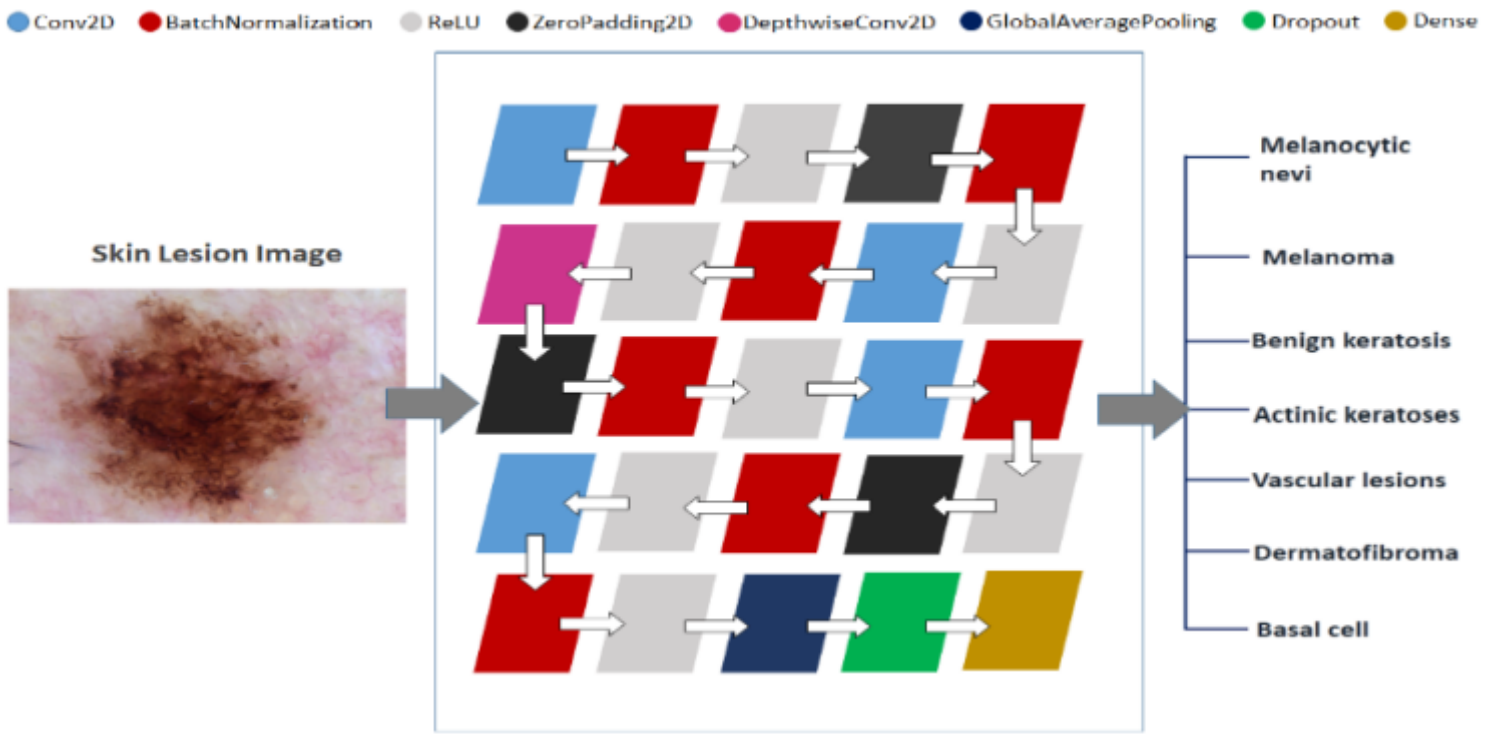

**Fig. 2.** MobileNet Architecture used in the current study for the classification of skin lesion image among seven skin cancer types.

## 2.3 Data augmentation

HAM10000 dataset has an unbalance distribution of images among the seven classes. Data Augmentation [42] brings an opportunity to rebalance the classes in the dataset, alleviating other minority classes. Data Augmentation is an effective means to expand the size of training data by randomly modifying several parameters of training data images like rotation range, zoom range, horizontal and vertical flip, fill_mode, etc. [42]. We conducted data augmentation of minority classes in the dataset: Melanoma, Benign Keratosis, Basal Cell Carcinoma, Actinic Keratosis, vascular lesion, and dermatofibroma to generate approximately 6000 images in each class giving a total of 38,569 images in the training set.

## 2.4 Training algorithm

The MobileNet model is ideal for mobile and embedded vision applications as they have lightweight DNN architecture [37]. We used MobileNet convolutional neural network [37] pretrained on 12,80,000 images containing 1,000 object classes from the 2014 ImageNet Challenge [31]. The 25 layered MobileNet architecture was constructed for the current study, which employs four Conv2D layers, seven BatchNormalization layers, seven ReLU layers, three ZeroPadding2D layers, and single DepthwiseConv2D, GlobalAveragePooling, Dropout, and Dense layers as shown in Figure 2. The training of the model was done on a training set of 38,569 images using Transfer Learning [39] with batch size and epochs as 10 and 50 respectively. The Categorical Crossentropy loss function, Adam optimizer, and metric function Accuracy, Top2 accuracy, and Top3 accuracy were used to evaluate MobileNet model performance.

### 2.5 Evaluation metrics

The overall performance of the model was evaluated with several evaluation metrics: Accuracy, Micro Average of Precision (MAP), Micro Average of Recall (MAR), and Micro Average of F1-score (MAF). The weighted average for Recall, Precision, and F1-score was evaluated by using the following mathematical expressions.

$$Accuracy = \frac{(TP+TN)}{(TP+TN+FP+FN)} \qquad (1)$$

$$Precision = \frac{TP}{(TP+FP)} \qquad (2)$$

$$Recall = \frac{TP}{(TP+FN)} \qquad (3)$$

$$F1 - score = 2\left(\frac{Precision * Recall}{Precision + Recall}\right) \qquad (4)$$

## 3 Results

The calculations were performed on Kaggle kernel having 4 CPU cores with 17 GB RAM and 2 CPU cores with 14GB RAM [43]. Model Evaluation was performed by calculating categorical accuracy, top2 accuracy, top3 accuracy, classification report, and confusion matrix. Further, the loss and accuracy curves were plotted to validate the model's performance for the optimization and prediction phase.

### 3.1 Data-set analysis

The important observations recorded during the data analysis of the HAM10000 dataset are shown in Figure 3 (i) Actinic Keratosis, Basal cell carcinoma, Dermatofibroma, and Melanocytic nevi are not much prevalent below the age of 20 years. Whereas Melanoma and Vascular lesions can occur at any stage of life. (ii) The peak age for skin cancer is found at 45 years, while they are more common between the age of 30 to 70. (iii) Back, Lower Extremity, Trunk, Upper Extremity and Abdomen are heavily compromised regions of skin cancer.

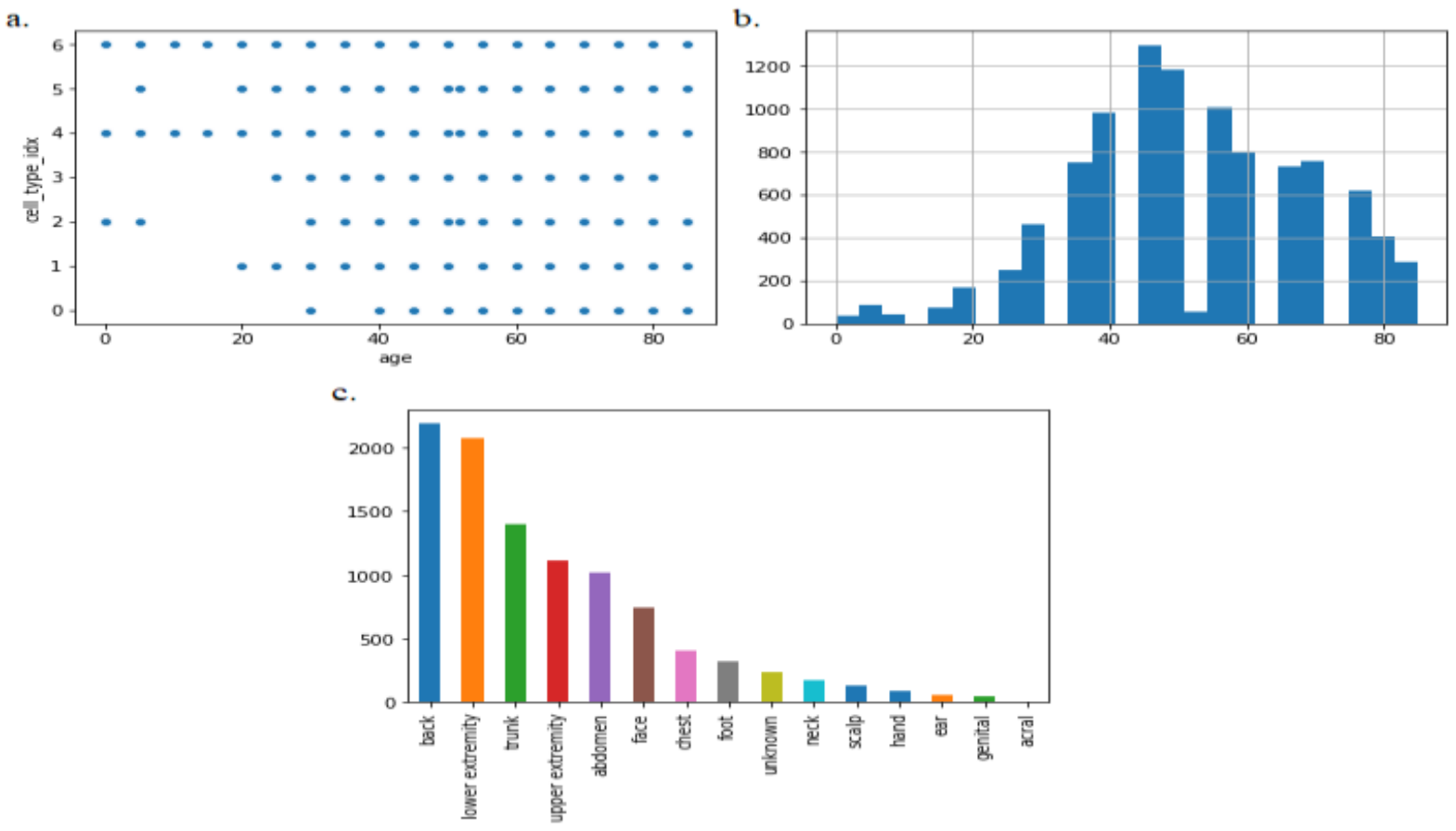

**Fig. 3.** Exploratory Data Analysis performed on the HAM10000 dataset (a) Comparative study of Skin cancer type on the y-axis with respect to age on the x-axis: The seven classes of the study represent 0,1,2,3,4,5 and 6 respectively on the y-axis. (b) Comparison of a number of cases of skin cancer on the y-axis with respect to age on the x-axis. (c) The number of skin cancer cases on the y-axis with respect to the location of skin cancer on the human body on the x-axis.

## 3.2 Model Validation

The validation of the model was conducted on 938 unknown sample images from the validation set. We evaluated micro and weighted average for precision, recall, and f1-score to evaluate the MobileNet model performance on unknown images of the validation set. The Weighted Average of 89%, 83%, 83%, and Micro Average of 83%, 83%, 83%, was recorded for Precision, Recall, and F1-score. The MobileNet model shows best precision, recall, and f1-score value for Melanocytic Nevi. The Multi-Class Classification Report showing Micro Average and Weighted Average for Precision, Recall, and F1-Score are represented in Table 1.

**Table 1.** Multi-Class Classification Report showing Micro Average and Weighted Average for Precision, Recall and F1-Score

| Classes | Precision | Recall | F1-Score |
|---|---|---|---|
| Actinic Keratosis | 0.36 | 0.38 | 0.37 |
| Basal Cell Carcinoma | 0.55 | 0.87 | 0.68 |
| Benign Keratosis | 1.00 | 0.13 | 0.24 |
| Dermatofibroma | 0.21 | 0.50 | 0.30 |
| Melanoma | 0.28 | 0.69 | 0.40 |
| Melanocytic Nevi | 0.95 | 0.93 | 0.94 |
| Vascular Lesions | 0.73 | 0.73 | 0.73 |
| **Micro Average** | **0.83** | **0.83** | **0.83** |
| **Weighted Average** | **0.89** | **0.83** | **0.83** |

The comparison of the current study with other related previous work is represented in Table 2. The majority of previous work is done on two or three classes, and their accuracies and recall vary between approximately 66 percent to 81 percent and 60 percent to 76 percent, respectively. In the study [21], they reported 48.9 percent and 55.4 percent classification accuracy evaluated for nine classes using CNN models. In the Study [34], classification accuracy for ten classes using Multi-track CNN was reported to be 75.1 percent. Also, in the study [35] they reported accuracy as 70 percent, 76 percent, 74 percent, and 67 percent for seven classes using InceptionResnetV2, PNASNet-5-Large, SENet154, and InceptionV4, respectively. In this study, we achieved categorical accuracy of 83.15 percent, top2 accuracy of 91.36 percent, top3 accuracy of 95.3 percent, and recall of 83 percent using MobileNet. Our seven-way skin cancer classification method has performed better than previously proposed computer-aided diagnosis systems in terms of both accuracies and recall. Additionally, the proposed method is more efficient considering the faster processing capability and lightweight architecture of MobileNet.

### 3.3 Confusion Matrix

The confusion matrix for our model was evaluated for seven classes. Each element of confusion matrix shows the comparison between the True label and Predicted label for each image in the validation set. Our model showed the best result for Melanocytic nevi by making a correct prediction for 696 images out of 751. Basal cell carcinoma and Melanoma were correctly determined for 26 images out of 30 and for 27 images out of 39, respectively. The diagnosis of Benign keratosis was most challenging due to their similar appearance with Melanoma and Melanocytic nevi. Only ten correct predictions were recorded for Benign keratosis.

**Table 2.** Comparison results of the Current Study with previous related work, * we have converted the recall and accuracy in percentage to compare them with the current study.

<table>
<tr><th>Source</th><th>Year</th><th>Classifier</th><th>No. of Classes</th><th>Accuracy %</th></tr>
<tr><td>[34]</td><td>2016</td><td>Multi-tract CNN</td><td>Ten</td><td>*75.1</td></tr>
<tr><td rowspan="2">[21]</td><td rowspan="2">2017</td><td>CNN<br>CNN-PA</td><td>Three</td><td>69.4<br>72.1</td></tr>
<tr><td>CNN<br>CNN-PA</td><td>Nine</td><td>48.9<br>55.4</td></tr>
<tr><td>[35]</td><td>2019</td><td>InceptionResnetV2<br>PNASNet-5-Large<br>SENet154<br>InceptionV4</td><td>Seven</td><td>70.0<br>76.0<br>74.0<br>67.0</td></tr>
<tr><td></td><td>2019</td><td>Current Study</td><td>Seven</td><td>83.15 (cat)<br>91.36 (top2)<br>95.84 (top3)</td></tr>
</table>

### 3.4 Loss and accuracy curves

In order to examine learning, generalizing and performance of the model, we computed training-validation loss curve (Figure 4(a) and training-validation accuracy curves for categorical (Figure 4(b)), top2 (Figure 4(c)) and top3 (Figure 4(d)) accuracies. The model shows a good learning rate as the training accuracy increase with the number of iterations along with symmetric downward sloping of the training loss curve. The small gap between training and validation curves represents a good-fit, showing model can generalize well on unknown images.

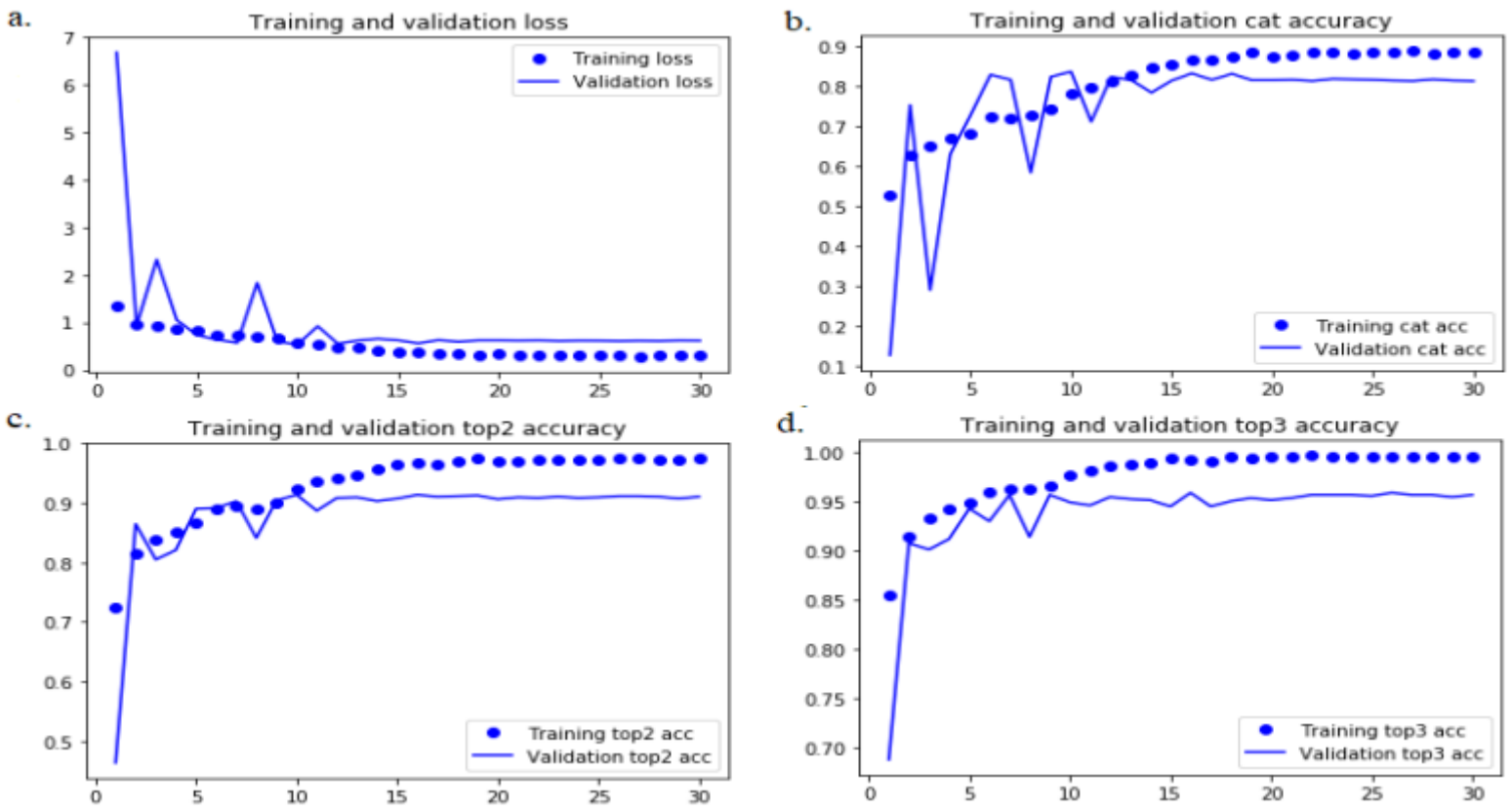

**Fig. 4.** Skin cancer classification performance curves of MobileNet model (a) Training and Validation loss (b) Training and Validation categorical accuracy (c) Training and validation top2 accuracy (d) Training and Validation top3 accuracy

We have developed a web application to provide an effective automated online tool for the multi-class classification of dermoscopy skin lesion images. This web application is available for social use at https://saketchaturvedi.github.io

# 4 Conclusion

The skin cancer incidences are intensifying over the past decades; the need of an hour is to move towards an efficient and robust automated skin cancer classification system, which can provide highly accurate and speedy predictions. In this study, we demonstrated the effectiveness of deep learning in automated dermoscopic multi-class skin cancer classification with the MobileNet model trained on a total of 38,569 dermoscopy images from HAM10000 dataset. We matched the performance of expert dermatologists across seven diagnostic tasks with an overall accuracy of 83.1% for seven classes in the dataset, whereas top2 and top3 accuracy of 91.36% and 95.34%, respectively. Also, the weighted average of precision, the weighted average of recall, and the weighted average of f1-score were found to be 89%, 83%, and 83%, respectively. We conclude that MobileNet model can be used to develop an efficient real-time computer-aided system for automated medical diagnosis systems. As compared to previously proposed models the MobileNet model has shown accurate and robust performance in addition to its faster and lightweight architecture.

The future work may deal with the utilization of patient's personalized data such as genes, age, color in addition to the current study for skin cancer diagnosis. This additional feature can be advantageous to develop personalized computer-aided systems for the diagnosis of skin cancers.